%% file: CONF_0lepton2011.tex
\RequirePackage{lineno}
\documentclass[final,5p,times,twocolumn]{elsarticle}

\usepackage{atlasphysics}
\usepackage{subfigure}
\usepackage{mathrsfs}
\usepackage{graphicx}
\usepackage{hyperref}
\usepackage{multirow}
\usepackage{rotating}
\usepackage{url}
\usepackage{textpos}
\usepackage{color}
\usepackage{epstopdf}
\usepackage{amsbsy}

\makeatletter
\g@addto@macro\bfseries{\boldmath}
\makeatother

\newcommand{\meff}{\ensuremath{m_{\mathrm{eff}}}}

\def\lsim{\mathrel{\rlap{\lower4pt\hbox{\hskip1pt$\sim$}}
    \raise1pt\hbox{$<$}}}                
\def\gsim{\mathrel{\rlap{\lower4pt\hbox{\hskip1pt$\sim$}}
    \raise1pt\hbox{$>$}}}                
\makeatletter
\renewcommand{\p@subfigure}{\arabic{figure}}
\renewcommand{\thesubfigure}{\alph{subfigure}}
\renewcommand{\@thesubfigure}{(\thesubfigure)}
\makeatother

\newcommand{\mytabref}[1]{Table~\ref{#1}}
\newcommand{\myfigref}[1]{Figure~\ref{#1}}

\newcommand{\anactualptmissvecwithapinit}{{\vec{P}}_\mathrm{T}^\mathrm{\; miss}}

\newcommand{\ourvecptmiss}{\ensuremath{\anactualptmissvecwithapinit}}
\newcommand{\ourmagptmiss}{\MET}

\newcommand{\ourdeltaphishort}{\Delta\phi}
\newcommand{\ourdeltaphifull}{\ourdeltaphishort(\textrm{jet},\ourvecptmiss)_\mathrm{min}}

\newcommand{\ph}{\phantom{0}}

\newcommand{\alpgen}{{\tt ALPGEN}}

\newcommand{\prospino}{{\tt PROSPINO}}
\newcommand{\herwig}{{\tt HERWIG}}
\newcommand{\pythia}{{\tt PYTHIA}}
\newcommand{\mcatnlo}{{\tt MC@NLO}}

\newcommand{\isajet}{{\tt ISAJET}}

\newcommand{\isasusy}{{\tt ISASUSY}}
\newcommand{\jimmy}{{\tt JIMMY}}

\newcommand{\cteqVER}[1]{{\tt CTEQ#1}}

\newcommand{\ourintlumi}{{$1.04~\ifb$}}
\newcommand{\ourpt}{p_\mathrm{T}}
\newcommand{\mSUGRA}{MSUGRA/CMSSM}

\newcommand{\degenerateMassLimitSqGlPlane}{1075}
\newcommand{\degenerateMassLimitSugraPlane}{950}
\newcommand{\absolteGluinoMassLimitSqGlPlane}{700}
\newcommand{\absolteSquarkMassLimitSqGlPlane}{875}

\begin{document}
\begin{frontmatter}
\title{Search for squarks and gluinos using final states with jets and missing transverse momentum with the ATLAS detector in $\sqrt{s}=7\TeV$ proton-proton collisions}
\author{The ATLAS Collaboration}

\begin{abstract}
\begin{textblock}{6}[0,1](0,-3){\small CERN-PH-EP-2011-145, Submitted to Physics Letters B}
\end{textblock}
A search for squarks and gluinos in events containing jets, missing
transverse momentum and no electrons or muons is presented. 
The data were recorded in 2011 by the ATLAS experiment in $\sqrt{s}=7\TeV$ proton-proton collisions at the Large Hadron Collider.
No excess above the Standard Model background expectation is observed in \ourintlumi{} of data.
Gluino and squark masses below \absolteGluinoMassLimitSqGlPlane~\GeV{} and \absolteSquarkMassLimitSqGlPlane~\GeV{} respectively are excluded at the 95\% confidence level   in simplified models containing only squarks of the first two generations, a gluino octet and a massless neutralino. The exclusion limit increases to \degenerateMassLimitSqGlPlane~\GeV{} for squarks and gluinos of equal mass. In \mSUGRA{} models with $\tan\beta=10$, $A_0=0$ and $\mu> 0$, squarks and gluinos of equal mass are excluded for masses below \degenerateMassLimitSugraPlane~\GeV. 
These limits extend the region of supersymmetric parameter space excluded by previous measurements. 
\end{abstract}

\end{frontmatter}

\section{Introduction}

Many extensions of the Standard Model (SM) include heavy coloured
particles, some of which could be accessible at the Large Hadron Collider (LHC) \cite{lhc}.  The squarks and
gluinos of supersymmetric (SUSY) theories \cite{susy} are one class of such
particles.
This Letter presents a new ATLAS search for squarks
and gluinos in final states containing only jets and large missing transverse momentum.
This final state can be generated by a large number of $R$-parity conserving
models~\cite{r-parity} in which squarks, $\squark$, and gluinos, $\gluino$, can be produced in pairs \{$\gluino\gluino$, 
$\squark \squark$, $\squark
\gluino$\} and can decay via
$\squark \rightarrow q
\ninoone$
and $\gluino
\rightarrow q \bar{q} \ninoone$ to weakly interacting neutralinos, $\ninoone$, which escape the detector
unseen.
The analysis presented here is based on a purely hadronic selection; events with reconstructed electrons or muons are
vetoed to avoid overlap with a related ATLAS search \cite{Collaboration:2011hh}. 
This updated analysis uses \ourintlumi of data recorded in 2011 and extends the sensitivity of the previous search described in Ref.~\cite{daCosta:2011qk} by including final state topologies with at least four jets, rather than three as before. The statistical analysis benefits from an improved technique which uses a combined likelihood fit across all the control regions used to determine the background contributions, in order to take into account correlations among the measurements.
The search strategy is optimised for
maximum discovery reach in the $(m_{\gluino},m_{\squark})$-plane for a set
of simplified models in which all other supersymmetric particles (except for
the lightest neutralino) are assigned masses beyond the reach of the LHC.
Currently, the most stringent limits on squark and gluino masses are obtained 
at the LHC~\cite{Collaboration:2011hh,  daCosta:2011qk, Khachatryan:2011tk}.

\section{The ATLAS Detector and Data Samples}

The ATLAS detector~\cite{Aad:2008zzm} is a multipurpose particle
physics apparatus with a forward-backward symmetric cylindrical
geometry and nearly 4$\pi$ coverage in solid angle.\footnote{
ATLAS uses a right-handed coordinate system with its origin at the nominal
interaction point in the centre of the detector and the $z$-axis along the
beam pipe. Cylindrical coordinates $(r,\phi)$ are used in the transverse
plane, $\phi$ being the azimuthal angle around the beam pipe. The pseudorapidity $\eta$ is
defined in terms of the polar angle $\theta$ as $\eta=-\ln\tan(\theta/2)$.
}  The layout of the detector is
dominated by four superconducting magnet systems, which comprise a
thin solenoid surrounding the inner tracking detectors and three large
toroids supporting a large muon spectrometer.
The calorimeters are of particular importance to this analysis.  In the pseudorapidity
region $\left|\eta\right| < 3.2$, high-granularity liquid-argon (LAr)
electromagnetic (EM) sampling calorimeters are used.  A steel-scintillator tile
calorimeter provides hadronic coverage over
$\left|\eta\right| < 1.7$.  The end-cap and forward regions, spanning
$1.5 < \left|\eta\right| < 4.9$, are instrumented with LAr calorimetry
for both EM and hadronic measurements.

The data used in this analysis were collected in the first half of 2011 with the LHC operating at a centre-of-mass energy of 7~\TeV.
Application of beam, detector and data-quality requirements resulted in
a total integrated luminosity of $1.04 \pm 0.04$~fb$^{-1}$~\cite{Atlas_lumi}.
The main trigger required events to contain a leading jet with a transverse momentum ($\ourpt$), above $75 \GeV$ and missing transverse momentum above $45 \GeV$. The trigger used an energy scale calibrated for electromagnetic objects. The details of the 
trigger specifications varied throughout the data-taking period, partly as a consequence of the rapidly
increasing LHC luminosity. The efficiency of the trigger is $>$ 98 \% for events selected by the offline analysis. The average number of proton-proton interactions per bunch crossing in the data sample was approximately six.

\section{Object Reconstruction}

The requirements used to select jets and leptons (objects) are chosen to give sensitivity to a range of SUSY models. Jet candidates are reconstructed using the
anti-$k_t$ jet clustering algorithm~\cite{Cacciari:2008gp,Cacciari:2005hq} with a
distance parameter of $0.4$. The inputs to this algorithm are three-dimensional clusters
of calorimeter cells \cite{larg-pub} seeded by those with energy significantly above
the measured noise. Jet momenta are constructed by performing a
four-vector sum over these cell clusters, treating each as an
$(E,\vec{p})$ four-vector with zero mass.  These jets are corrected for
the effects of calorimeter non-compensation and inhomogeneities by
using $\ourpt$ and $\eta$-dependent calibration factors
based on Monte Carlo (MC) and validated with extensive
test-beam and collision-data studies \cite{atlas-jetcal}. Furthermore, the reconstructed jet is modified such that the jet
direction points to the primary vertex, defined as the vertex with the highest summed track $\pt^2$, instead of the geometrical centre of the
ATLAS detector. Only jet candidates with corrected transverse momenta $\ourpt > 20$
\GeV{}
are subsequently retained.
For 84\% of the data used, a temporary electronics failure in the LAr barrel calorimeter created a dead region in the second and third longitudinal layers, approximately $1.4\times 0.2$ in $\Delta\eta\times\Delta\phi$, in which on average 30\% of the incident jet energy is lost. The impact on the reconstruction efficiency for $\pt$ $>$ 20 GeV jets is found to be negligible. If any of the four leading jets 
fall into this region the event is rejected, causing a loss of signal acceptance which is smaller than 15\% for the models considered here.

Electron candidates are required to have $\ourpt > 20~\GeV$, have $|\eta| <
2.47$, and pass the `medium'  shower shape and track selection criteria of
Ref.~\cite{Aad:2010yt}.
Muon candidates \cite{Aad:2010yt} are required to have $\ourpt > 10$ \GeV{} and $|\eta| <
2.4$.  
Since no use is made of tau-lepton candidates in this analysis, in the following the term lepton will refer only to electrons and muons.

The measurement of the missing transverse momentum two-dimensional vector
$\ourvecptmiss$ (and its magnitude $\ourmagptmiss$) is then based on
the transverse momenta of all electron and muon candidates, all jets which are not also electron candidates, and all
calorimeter clusters with $|\eta|<4.5$ not associated to such objects.  

Following the steps above, overlaps between candidate jets with $|\eta|<2.8$ and leptons are resolved using the method of Ref.~\cite{atlas-csc} as follows.
First, any such jet candidate lying within a distance $\Delta R = \sqrt{(\Delta \eta)^2 +
(\Delta \phi)^2}
=0.2$ of an electron is discarded:
then any electron or muon candidate remaining within a distance
$\Delta R =0.4$ of any surviving jet candidate is discarded. 
Next, all jet candidates with $|\eta|>2.8$ are discarded. 
Thereafter, the electron, muon and jet candidates surviving this procedure are considered as
``reconstructed'', and the term ``candidate'' is dropped.

\section{Event Selection}\label{sec:eventselection}

Following the object reconstruction described above, events are
discarded if they contain any electrons or muons
with $\pt$ $>$ 20 GeV, or any jets
failing quality selection criteria designed to suppress detector noise and
non-collision backgrounds
(see e.g. Ref.~\cite{atlas-jet-clean}). 

These selections include a veto on leading jets (with $\ourpt > 100$ \GeV{} and $|\eta| <2$)
which have a low fraction ($< 0.05$) 
of their 
$\ourpt$ 
carried by charged tracks, 
and a requirement that the leading jets all have consistent timing information from the calorimeters.  Events are also rejected if the reconstructed primary
vertex is associated with fewer than five tracks.

In order to achieve maximal reach over the
$(m_{\gluino},m_{\squark})$-plane, five signal regions are defined. Squarks typically generate at least one jet in their decays, for instance through $\squark \rightarrow q
\ninoone$, while gluinos typically generate at least two, for instance through $\gluino\rightarrow q \bar{q} \ninoone$. Processes contributing to $\squark\squark$, $\squark\gluino$ and $\gluino\gluino$ final states therefore lead to events containing at least two, three or four jets, respectively. Cascade decays of heavy particles tend to increase the final state multiplicity. 
Four signal regions characterized by increasing jet multiplicity requirements are therefore defined as shown in \mytabref{tab:signal_regions}, with the leading jet having 
$\ourpt>130$~GeV, and other jets $\ourpt> 40$~GeV.  
The effective mass, \meff, is calculated as the sum of $\ourmagptmiss$ and the magnitudes of the transverse momenta of the two, three or four highest $\pT$ jets used to define the signal region.
Two four-jet signal regions are defined requiring $\meff>500$~GeV (optimised for small mass differences between SUSY mass states) and $\meff>1000$~GeV (optimised for higher mass differences). 
In addition, a fifth `high mass' signal region is derived from the four-jet sample, with more stringent requirements on the $\ourpt$ of the non-leading jets ($ >80$~GeV) 
and on $\meff ~(>1100$~GeV), in order to give maximal reach in the SUSY mass spectrum. For this latter signal region the transverse momenta of all jets with $\pt>40$~GeV are used to compute $\meff$.
In \mytabref{tab:signal_regions}, $\ourdeltaphifull$ is the smallest of the
azimuthal separations between $\ourvecptmiss$ and jets with $\ourpt>40~\GeV$ (all reconstructed jets up to a maximum of three, in descending order of $\ourpt$). 
Requirements on $\ourdeltaphifull$ and $\ourmagptmiss/\meff$ are designed to reduce the background from multi-jet processes.

\newcommand\ourtwidth{2.0cm}
\newcommand\yourtwidth{1.5cm}
\newcommand\errorSTAT{(m)}
\newcommand\errorUNCORR{[u]}
\newcommand\errorUNCORRANDJES{[u+j]}
\newcommand\errorJES{[j]}
\newcommand\errorJER{(r)}
\newcommand\errorLUMI{[{$\mathcal L$}]}
\newcommand\errorACC{(a)}
\newcommand\errorLE{($\epsilon$)}
\newcommand\errorCR{(c)}
\newcommand\errorEXT{(x)}
\newcommand\errorCREXT{(c+x)}

\begin{table}
  \begin{center}\renewcommand\arraystretch{1.2}
\renewcommand\tabcolsep{1.2mm}
    \begin{tabular}{| l l | c|c|c|c|}
 \hline \hline
 & Signal Region &$\geq$ 2-jet & $\geq$ 3-jet & $\geq$ 4-jet & High mass \\
\hline
\hline
& $\ourmagptmiss$            & $>130$ &  $>130$ &  $>130$  & $>130$ \\ 
& Leading jet $\ourpt$       & $>130$ &  $>130$ &  $>130$  & $>130$ \\ 
\hline
& Second jet      $\ourpt$   & $>40$ &  $>40$ &  $>40$ & $>80$\\ 
& Third jet      $\ourpt$       &  --         &  $>40$ &  $>40$ & $>80$\\ 
& Fourth jet      $\ourpt$       &  --        &     --       &  $>40$ & $>80$\\ 
\hline
&$ \ourdeltaphishort(\textrm{jet},\ourvecptmiss)_\mathrm{min}$ & $>0.4$ & $>0.4$  & $>0.4$ & $>0.4$   \\ 
& $\ourmagptmiss/\meff$     & $>0.3$     & $>0.25$ & $> 0.25$ & $>0.2$ \\ 
& $\meff$   & $>1000$  &  $>1000$ & $>500/1000$ & $>1100$ \\ 
\hline \hline
\end{tabular}
\caption{Criteria for admission to each of the five overlapping signal regions ($\meff$, $\ourmagptmiss$ and $\ourpt$ in GeV).
 All variables are defined in Section \ref{sec:eventselection}. The $\meff$ is defined with a variable number of jets, appropriate to each signal region. In the high mass selection, all jets with $\ourpt>40$~GeV are used to compute the $\meff$ value used in the final cut, but the $\ourmagptmiss$ definition is unchanged. The $\ourdeltaphishort$ cut is only applied up to the third leading jet. }
    \label{tab:signal_regions}
  \end{center}
\end{table}

\section{Backgrounds, Simulation and Normalisation}

Standard Model background processes contribute to the event counts in
the signal regions. The dominant sources are: $W+$jets, $Z+$jets, top
pair, single top, and multi-jet    
production.
Non-collision backgrounds have been found to be negligible.
The majority of the $W$+jets background is composed 
of $W\rightarrow \tau\nu$ events, or $W\rightarrow e\nu, \mu\nu$ events in which  
no electron or muon candidate is reconstructed.
The largest part of the $Z$+jets background comes from the irreducible
component in which $Z\rightarrow\nu\bar\nu$ decays generate large $\ourmagptmiss$. 
Hadronic $\tau$ decays in $t \bar t \rightarrow b \bar b \tau \nu q q $ and single top events
can also generate large $\ourmagptmiss$ and pass the jet
and lepton requirements at a non-negligible rate.
The multi-jet background in the signal regions is caused by
misreconstruction of jet energies in the calorimeters leading to apparent
missing transverse momentum, as well as by neutrino production in semileptonic decays of heavy quarks. Extensive validation of the MC simulation against data has been performed for each of
these background sources and for a wide variety of control
regions.

In order to estimate the backgrounds in a consistent fashion, five control regions (CRs) are defined for each of the five signal regions (SRs), giving 25 CRs in total. The orthogonal CR event selections are designed to provide uncorrelated data samples enriched in particular background sources. Each ensemble of one SR and five CRs constitutes a different `channel' of the analysis. The CR selections are optimised to maintain adequate statistical weight, while minimising as far as possible the systematic uncertainties arising from extrapolation to the SR. The purities of the CRs for the main background processes in which they are enriched exceed 50\% in all cases.

For each channel, measurements in the CRs are used to derive background expectations in the SR through the use of `transfer factors' equivalent to the ratios of expected event counts in the CRs and SR, derived independently of the data observations in the CR and SR.
Some uncertainties, such as those arising in MC simulation from the jet energy scale and physics modelling, are reduced in the transfer factors. The combined likelihood fit across all control regions ensures that the background estimates are consistent for all processes, taking into account contamination of the CRs by multiple SM processes.

The likelihood function is built by the Poisson probability density function (pdf) describing the SR and the CRs and a pdf describing the systematic uncertainties:
$$
  L({\boldsymbol n}|\mu,{\boldsymbol b},{\boldsymbol\theta})=P_{\rm SR} \; \times \; P_{\rm WR}  \;\times\;
 P_{\rm TR}  \;\times\;  P_{\rm ZRa}  \;\times\;  P_{\rm ZRb}  \;\times\;  P_{\rm QR} \;\times\; C_{\rm Syst}  .
$$

The mean of the Poisson pdfs in the CRs are defined as
$$
\lambda_{i}(\mu,{\bf b,\boldsymbol\theta}) = \mu \cdot
C_{\rm{SR}\rightarrow i\rm{R}}({\boldsymbol\theta}) + \sum_{j}
C_{j\rm{R}\rightarrow i\rm{R}}({\boldsymbol\theta})\cdot b_{j\rm{R}}
$$
where the index $j$ runs over the background control regions.  
$\mu$ is the signal strength, $b_{jR}$ is the background j in Region R and $C_{jR->SR}$ the transfer factor of process j from region R to the SR.
The terms $C_{\rm{QR}\rightarrow \rm{QR}}$, $C_{\rm{WR}\rightarrow \rm{WR}}$,$C_{\rm{ZRa,b}\rightarrow \rm{ZRa,b}}$,
$C_{\rm{TR}\rightarrow \rm{TR}}$ are by construction all equal to $1$.
Since the fit is not over-constrained in CR2, CR3 and CR4, where there is a single estimate of the background, the fit output matches the observed number of events in these regions by construction. This is not the case in CR1a and CR1b which both estimate the same background process, and a best fit number is produced in these regions.

The transfer factors are obtained from a combination of data and MC inputs.
Those for multi-jet processes are estimated using a data-driven technique based upon the smearing of jets in a low $\ourmagptmiss$ data sample (`seed' events with $\met/\sqrt{\Sigma p_T({\rm jet})}<0.6$ GeV$^{1/2}$) with jet response functions tuned by comparison with multi-jet dominated data control regions \cite{daCosta:2011qk}. For the $Z$+jets, $W$+jets and top quark processes they are derived from MC. For each channel a likelihood fit is performed to the observed event counts in the five CRs, taking into account correlations in the systematic uncertainties in the transfer factors. 

The irreducible background from $Z(\nu\bar{\nu})$+jets events is estimated using control regions enriched in related processes with similar kinematics: events with isolated photons and jets \cite{Ask:Zgamma} 
and events due to $Z(ee/\mu\mu)$+jets (control regions denoted `CR1a' and `CR1b' respectively). The reconstructed momentum of the photon or the lepton-pair system is added to $\ourvecptmiss$ to obtain an estimate of the $\ourmagptmiss$ observed in $Z(\nu\bar{\nu})$+jets events. The results from both control regions are found to be in good agreement, and both are used in the final fit. The small additional background contributions arising from $Z$ decays to misidentified charged leptons, and misidentified photon events, are estimated using the same control regions with appropriate transfer factors.

The background from multi-jet processes is determined using control regions (CR2) in which the cut on $\ourdeltaphifull$ is reversed and tightened: $\ourdeltaphifull<0.2$. This selects events in which $\ourvecptmiss$ is aligned with one of the three leading jets in the transverse plane. Such a topology is characteristic of events containing mismeasured jets, or neutrino emission from heavy flavour decays within jets. A separate control region is used to estimate the additional multi-jet background generated by events affected by the temporarily dead region in the barrel EM calorimeter; this result is added to the multi-jet background estimate obtained from CR2.

The background from $W(\ell\nu)$+jets production is estimated from samples of events with a lepton ($\ell$), $\MET>130$ GeV and a transverse mass of the $(\ell , \MET)$ system between 30 GeV and 100 GeV, i.e. consistent with the $W$ mass (control regions CR3). A veto against jets arising from $b$-quark decays, based on a tagging procedure exploiting both impact parameter and secondary vertex information, is applied to remove events containing top quarks. In this CR, leptons are treated as jets for the computation of the kinematic variables. 

The background from top quark production is estimated using the same selection as for $W(\ell\nu)$+jets events, but replacing the $b$-jet veto with a $b$-tag requirement (control regions CR4). This enhances the population of events containing top quark decays relative to that of direct $W$ production events. The resulting transfer factors include the contribution from events where both top quarks decay semi-leptonically, as well as events due to single top production.

MC simulation samples are used to develop the analysis, determine the transfer factors used to estimate the $W$+jets, $Z$+jets and top quark backgrounds, and assess the sensitivity to specific SUSY signal models.  Samples of multi-jet events from quantum-chromodynamic (QCD) processes are generated with \pythia~\cite{Sjostrand:2006za}, using the {\tt MRST2007LO*} modified leading-order parton distribution functions (PDFs)~\cite{Sherstnev:2007nd}. Production of top quark pairs is simulated with \mcatnlo~\cite{Frixione:2002ik,Frixione:2003ei} (with a top quark mass of $172.5$ \GeV) and the Next-to-Leading Order (NLO) PDF set {\tt CTEQ6.6}~\cite{Nadolsky:2008zw}. 
Single top production is also simulated
with \mcatnlo~\cite{Frixione:2005vw,Frixione:2008yi}.
Samples of $W$ and Z/$\gamma^*$ events with accompanying jets are generated
with \alpgen~\cite{Mangano:2002ea} and PDF set {\tt CTEQ6L1}~\cite{Pumplin:2002vw}. 
Fragmentation and hadronization for the \alpgen~and \mcatnlo~samples is
performed with \herwig~\cite{Corcella:2000bw,herwig65long}, using
\jimmy~\cite{Butterworth:1996zw} for the underlying event. SUSY signal samples are generated with {\tt HERWIG++}~\cite{Bahr:2008pv}, normalised using NLO cross sections determined with \prospino{}
\cite{Beenakker:1996ch}. The MC samples are produced using ATLAS 
parameter 
tunes \cite{MC10a} and are processed through a {\tt GEANT4}~\cite{Agostinelli:2002hh} based detector simulation \cite{:2010wqa}. Corrections are applied for small differences in 
reconstruction efficiencies, energy scales and resolutions between data and MC. Varying pile-up conditions as a function of the instantaneous luminosity are taken into account by reweighting the simulated events according to the mean number of interactions per bunch crossing observed in the data. Multi-jet MC samples, presented here in some figures for illustrative purposes only, are normalised to a sample of dijet events with $\ourdeltaphifull<0.4$. In all other cases the best available NLO or Next-to-NLO theoretical cross-section calculations were used.

\section{Systematic Uncertainties}\label{sec:systematics}

Systematic uncertainties arise from the use of the transfer factors relating observations in the control regions to background expectations in the signal regions, and from the modelling of the SUSY signal. For the transfer factors derived from MC, the primary common sources of systematic uncertainty are the jet energy scale and resolution, physics modelling and reconstruction performance in the presence of pile-up.
 
The jet energy scale uncertainty has been
measured from the complete 2010 data set using the procedure
described in Ref.~\cite{atlas-jetcal}.
It depends upon $\ourpt$, $\eta$ and proximity to adjacent jets, and on average amounts to around 4\%.
The jet energy resolution measured with 2010 data \cite{atlas-jer} is applied to the MC
jets, with the difference
between the re-calibrated and nominal MC resolution taken as the systematic uncertainty.
Additional contributions are added to both of these uncertainties to take into account of the impact of pile-up at the relatively high luminosity 
delivered by the LHC in the 2011 run. 
Both in-time pile-up, i.e. multiple collisions within the same bunch crossing, and out-of-time pile-up, which arises from the detector response to 
neighbouring bunch crossings, have effects on jet energy measurements. These were studied in detail as a function of the average number of 
collisions per bunch crossing and by comparing data recorded with 75 and 50~ns bunch spacing. A worsening in the jet energy resolution in 
the forward region is observed when moving from 75 to 50 ns operation; a systematic uncertainty of $0.07\times \ourpt$ is therefore applied to 
jets with $|\eta|>2.8$,  used for the $\ourmagptmiss$ calculation. The combined effects of in-time and out-of-time pile-up on the jet energy scale are accounted for by an additional 
conservative systematic uncertainty of up to 7\% depending on $|\eta|$ and $\ourpt$. All these uncertainties are propagated to the 
$\ourmagptmiss$ measurement. The impact of in-time pile-up on other aspects of the selection was also investigated and found to be negligible 
as expected given the high energies of the jets entering the signal samples.

The dominant modelling uncertainty in MC predictions for the signal region and control regions arises from the treatment of jet radiation,which affects the calculation of  $\meff$. In order to assess this uncertainty, the main backgrounds are estimated using alternative generators (\alpgen~rather than \mcatnlo~for $t\bar{t}$ production) or reduced jet multiplicity (\alpgen~processes with 0--4 partons instead of 0--5 partons for $W/Z$+jets production). The impact of renormalisation and factorisation scale variations and PDF uncertainties was also studied. Differences in the absolute expectations for the numbers of events in the SR and CR as high as 100\% are observed for specific processes; the impact on the ratios used in the transfer factors is, however, much smaller (differences $\lsim$40\%, channel dependent). 

Additional uncertainties considered, for specific processes, include those arising from photon and lepton trigger efficiency, reconstruction efficiency, energy scale and resolution (CR1a, CR1b, CR3 and CR4), $b$-tag/veto efficiency (CR3 and CR4), photon acceptance and backgrounds (CR1a) and the limited size of MC samples (all CRs). Uncertainties on the multi-jet transfer factors are dominated by the modelling of the non-Gaussian tails of the response function. Other sources, including the limited number of data events, and  uncertainties on the Gaussian part of the response functions, are also considered.

Systematic uncertainties on the expected SUSY signal are estimated by varying
the factorisation and renormalisation scales in \prospino{} between half
and twice their default values and by considering the PDF uncertainties provided by \cteqVER{6}.
Uncertainties are calculated for individual
production processes ($\squark\squark$, $\gluino\gluino$, and $\squark\gluino$) 
and are typically $\sim$35\% in the vicinity of the limits expected to be set by this analysis. Jet energy scale and resolution, and pile-up uncertainties on SUSY signal expectations are typically smaller than 30--40\%.

\section{Results, Interpretation and Limits}

\begin{figure*}
  \centering
  
  \includegraphics[width= 0.485\textwidth]{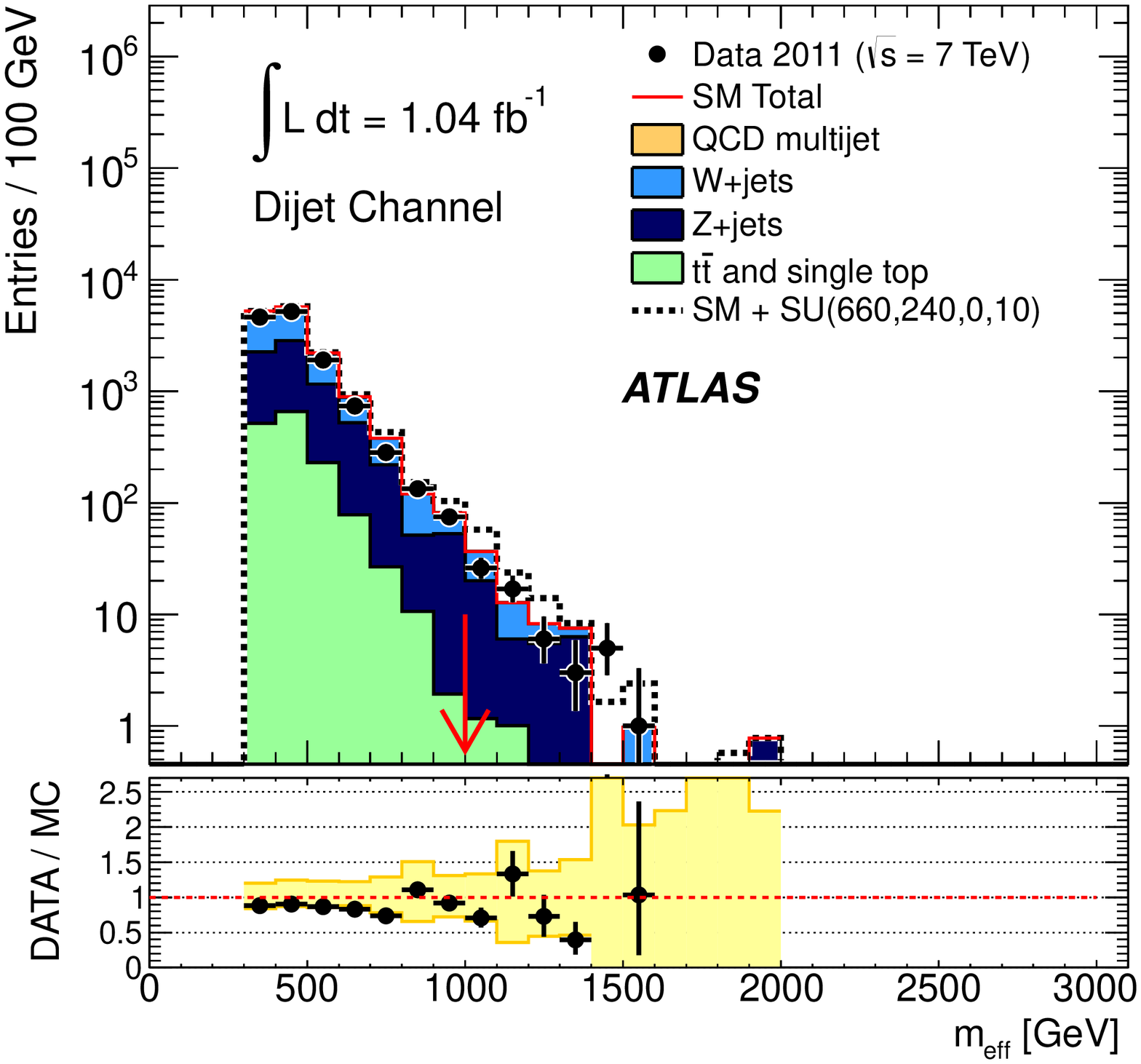}       
  \includegraphics[width= 0.485\textwidth]{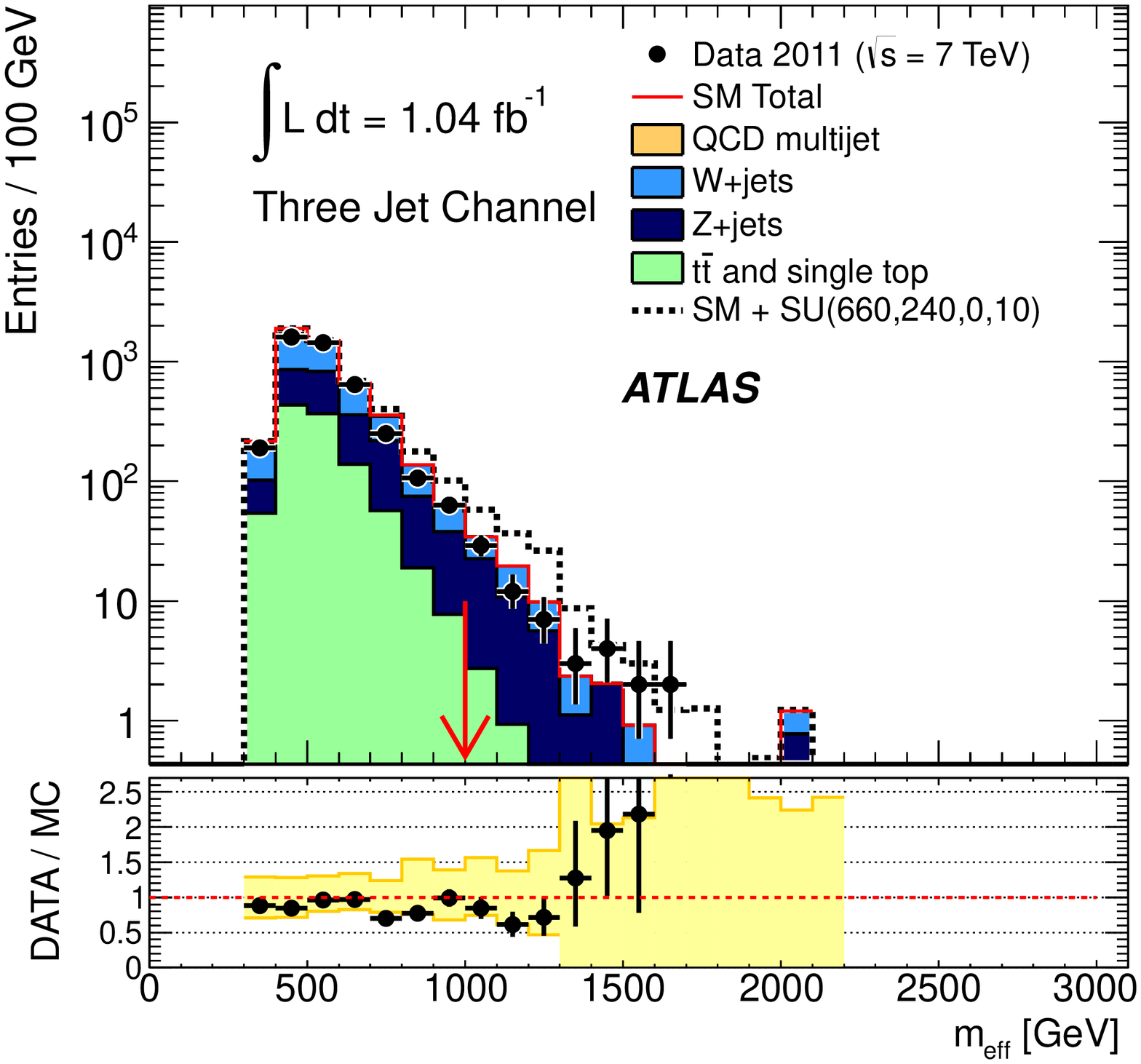}
  \includegraphics[width=0.485\textwidth]{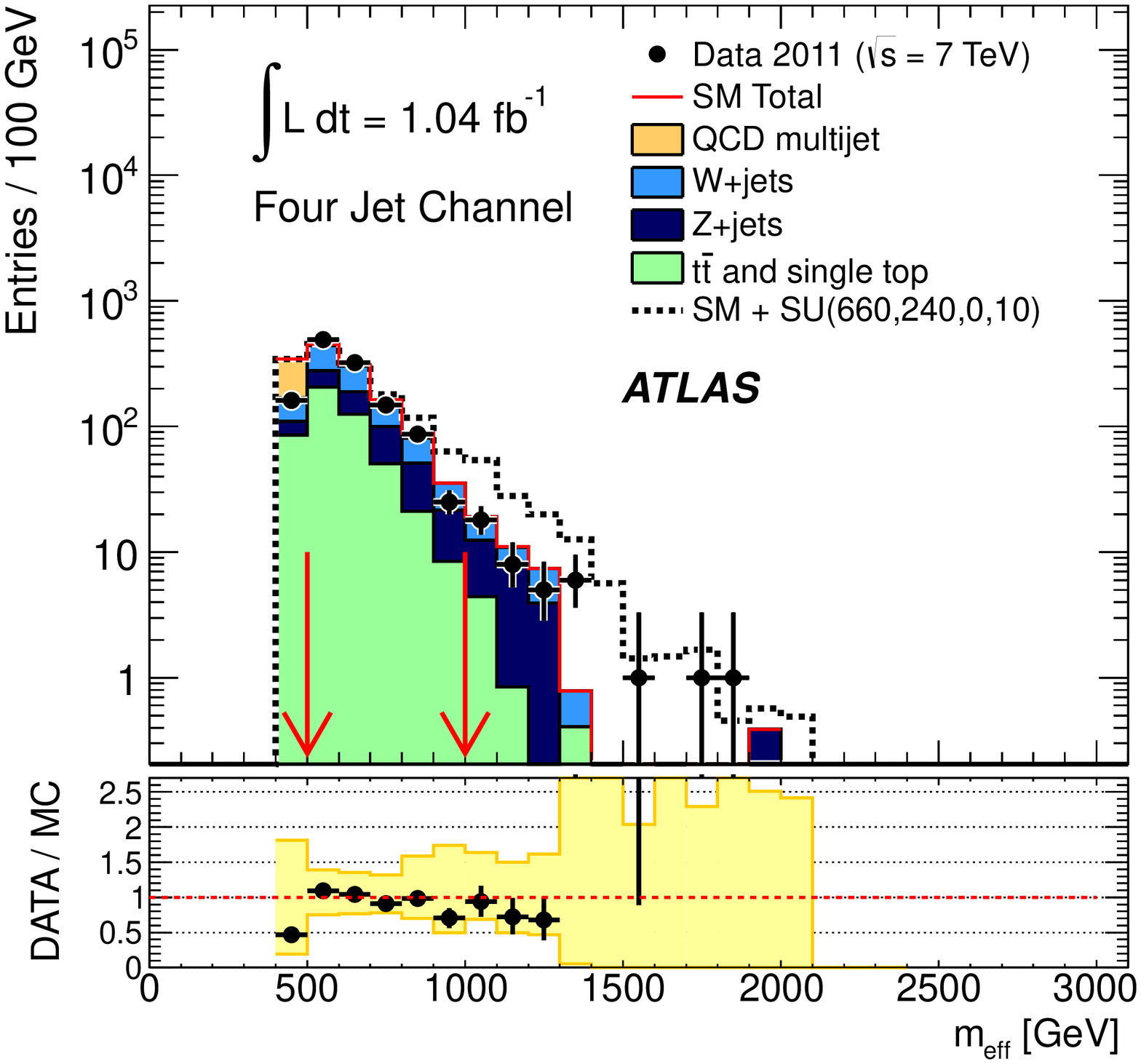}
   \includegraphics[width=0.485\textwidth]{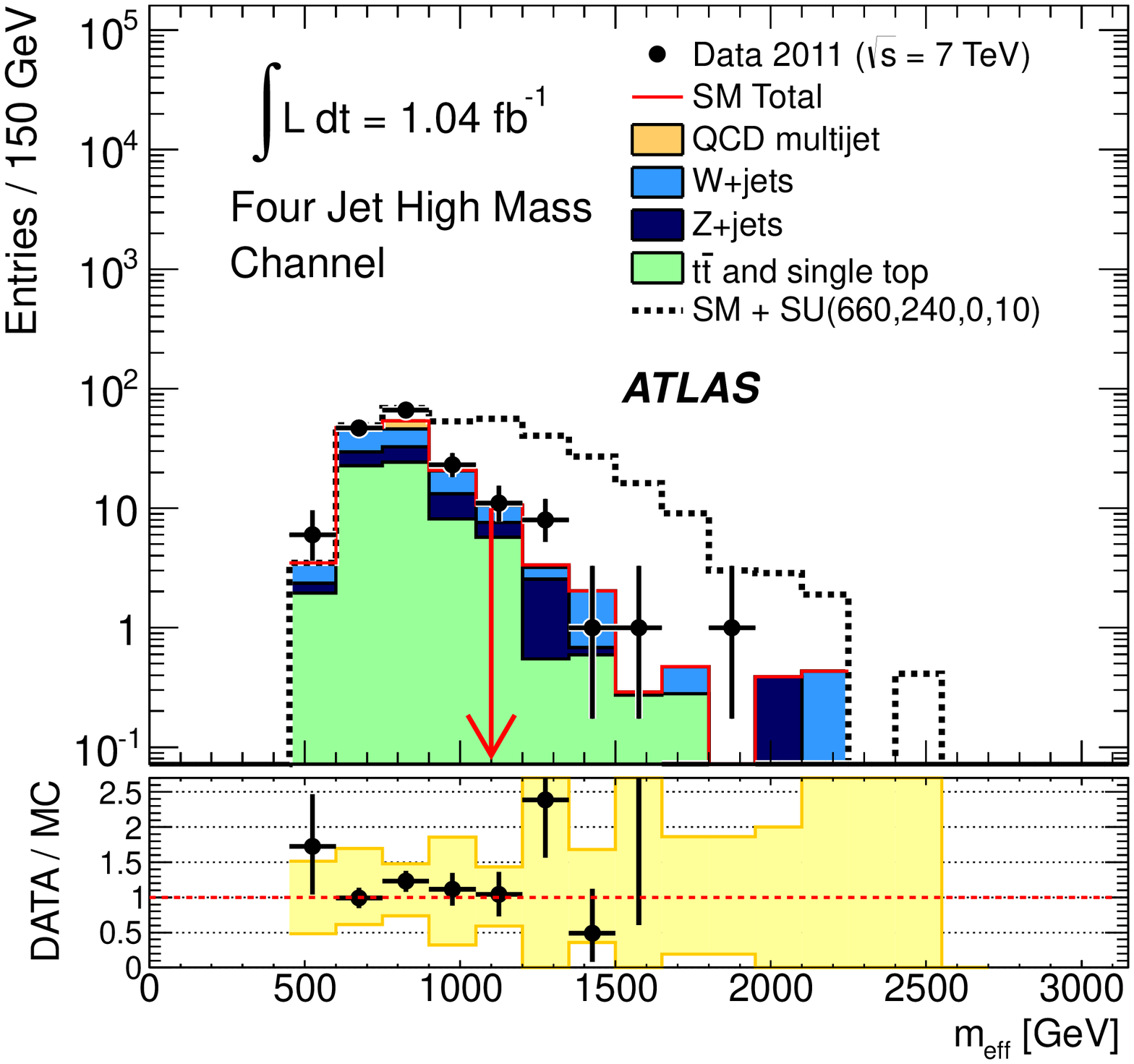}
  
  \caption{The observed \meff{} distributions in the signal regions for the $\ge 2$ jet channel (top left), the $\ge 3$ jet channel (top right) and the two $\ge 4$ jet channels (bottom left), and for the high mass  channel using the inclusive definition of \meff{} (bottom right), after all the selection criteria but the $\meff$ cut. 
These plots also show the expected SM contributions obtained from MC simulated samples prior to normalisation using the data-driven likelihood method described in the text. 
The red arrows indicate the lower bounds on $\meff$ used in the final signal region selections.
The expectation for a \mSUGRA{} reference point with  $m_0=660$~\GeV, $m_{1/2}=240$~\GeV, $A_0=0$, $\tan\beta=10$ and $\mu> 0$ is also shown.  
This reference point is also indicated by the star on \myfigref{fig:limitcombinedCLs}.  
Below each plot the ratio of the data to the SM expectation is provided.  Black vertical bars show the statistical uncertainty from the data, while the yellow band shows the size of the systematic uncertainties from the MC simulation.
  \label{fig:meff_data}}
\end{figure*}

The observed signal region $\meff$ distributions for each of the channels used in this analysis are shown in \myfigref{fig:meff_data}, together with MC background expectations prior to using the likelihood fitting procedure.  The number of observed data events and the number of SM events expected to enter each of the signal regions, determined using the likelihood fit, are shown in \mytabref{tab:SRevc}. The data are found to be in good agreement with the background expectation and no excess is observed. To illustrate the procedure, the inputs and outputs of the combined likelihood fit for the high mass channel are shown in \mytabref{tab:SREdetails}.

\input{table2}
\input{table3}

Data from the five channels are used to set the limits, taking the channel with the best expected limit at each point in parameter space. The limit for each channel is obtained by comparing the observed numbers of signal events with those expected from SM background plus SUSY signal processes, taking into account uncertainties in the expectation including those which are correlated between signal and background (for instance jet energy scale uncertainties). The impact of SUSY signal contamination of the control regions is taken into account by applying MC-derived model dependent correction factors $\sim0.97$--1.02 to the resulting exclusion significance values. The excluded regions are obtained using the CL$_{\rm s}$ prescription~\cite{clsread}. 
\par

\begin{figure*}
\begin{center}

\subfigure{\includegraphics[width=0.45\textwidth]{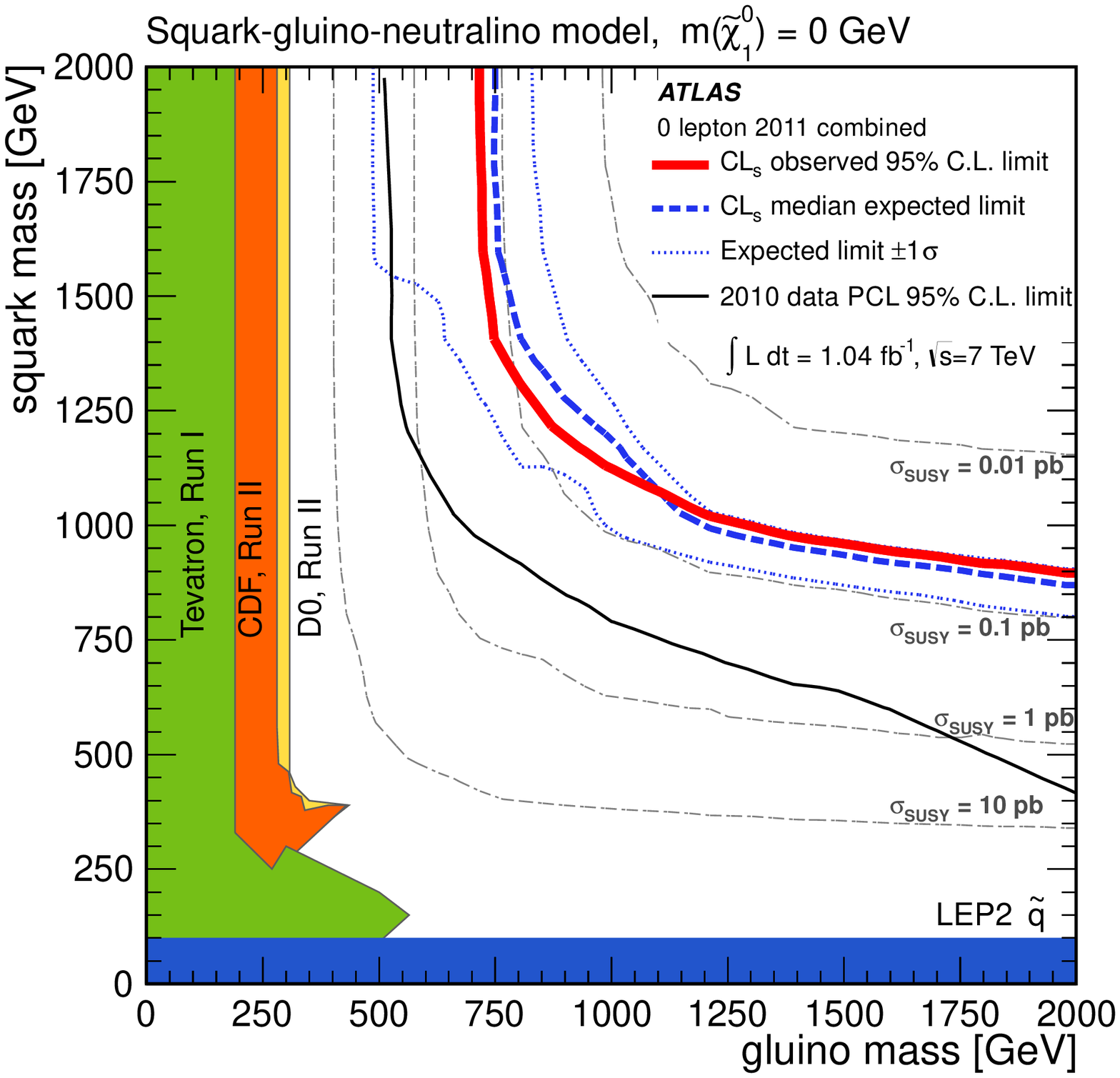}}
\subfigure{\includegraphics[width=0.45\textwidth]{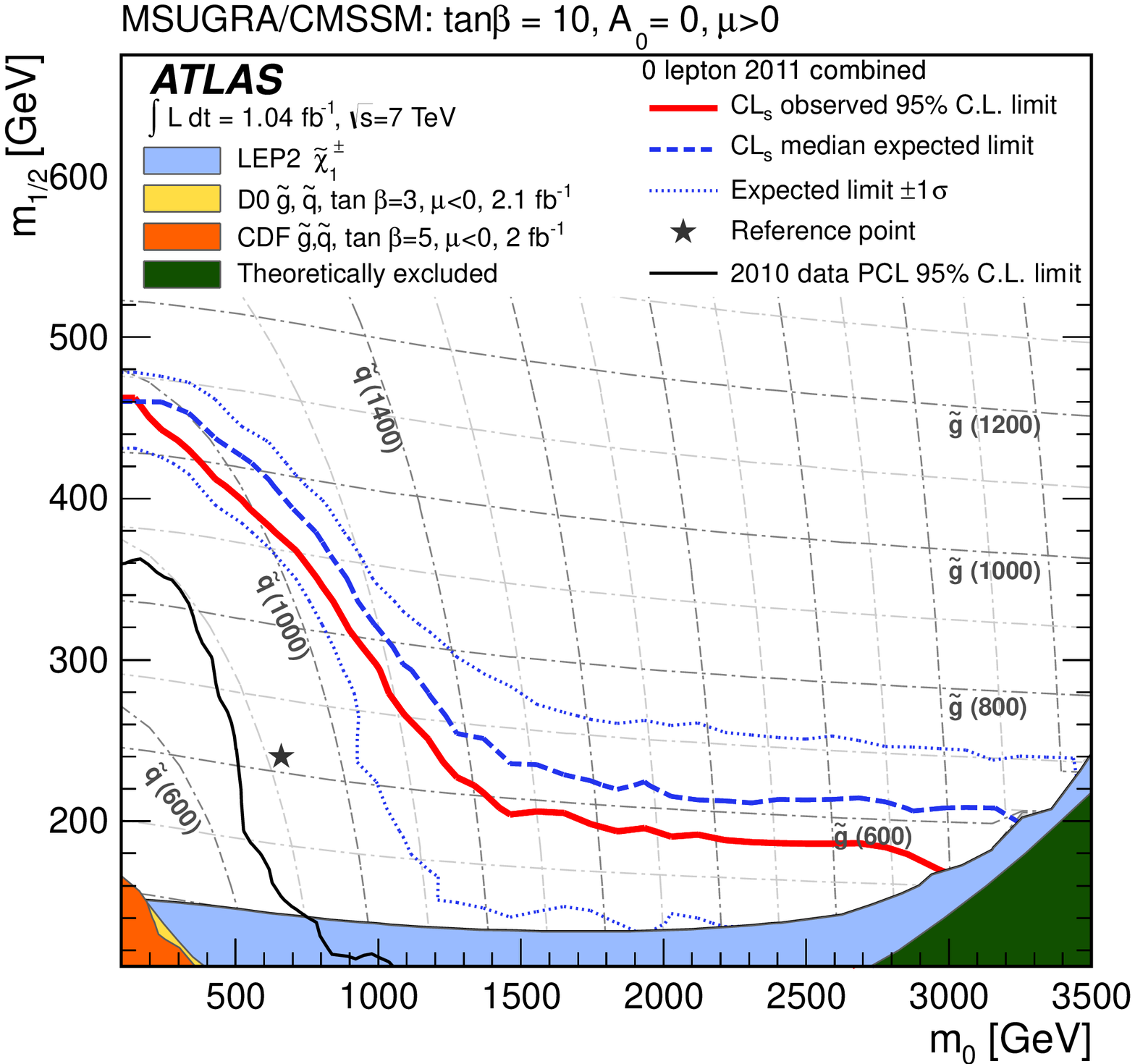}}

 \caption{Combined exclusion limits for simplified SUSY models with $m(\ninoone)=0$ (left) and MSUGRA/CMSSM models with $\tan\beta=10$, $A_0=0$ and $\mu>0$ (right). The combined limits are obtained by using the signal region which generates the best expected limit at each point in the parameter plane. The dashed-blue line corresponds to the 
median expected 95\% C.L. limit and the red line corresponds to the observed limit at 95\% C.L. The dotted blue lines correspond to the $\pm1\sigma$ variation in the expected limits. Also shown for comparison purposes in the figures are limits from the Tevatron~\cite{Abachi:1995ng, Affolder:2001tc, Aaltonen:2008rv, :2007ww} and LEP~\cite{Abdallah:2003xe,Achard:2003ge}
, although it should be noted that some of these limits were generated with different models or parameter choices (see legends). The previous published ATLAS limits from this analysis \cite{daCosta:2011qk} are also shown. The \mSUGRA{} reference point used in \myfigref{fig:meff_data} is indicated by the star in the right-hand figure.  
\label{fig:limitcombinedCLs}}
\end{center}
\end{figure*}

An interpretation of the results is presented in
\myfigref{fig:limitcombinedCLs}{} (left) as a 95\% confidence exclusion
region in the $(m_{\gluino},m_{\squark})$-plane for a simplified set of SUSY
models with $m(\ninoone)=0$.  In these models the gluino mass and the masses of the
squarks of the first two generations are set to
the values shown in the figure. All other supersymmetric particles,
including the squarks of the third generation, are decoupled by being given masses of 5~\TeV. The limits are reduced by decay chain kinematics if $m(\ninoone)$ is comparable to the squark or gluino mass.
{\isasusy{} from
\isajet{} \cite{Paige:2003mg} {v7.80} is used to calculate the
decay tables, and to guarantee consistent electroweak symmetry
breaking.}

The results are also interpreted in the
$\tan\beta=10$, $A_0=0$, $\mu>0$ slice of \mSUGRA\footnote{Five parameters are needed to specify a particular \mSUGRA\ model.  They are the universal scalar mass, $m_0$, the universal gaugino mass $m_{1/2}$, the universal trilinear scalar coupling, $A_0$, the ratio of the vacuum expectation values of the two Higgs fields, $\tan\beta$, and the sign of the higgsino mass parameter, $\mu>0$ or $ <0$. }~\cite{msugraandcmssm}
in \myfigref{fig:limitcombinedCLs}{} (right). These limits include the effects of the mass spectrum of the SUSY particles on their decay chains. In regions of parameter space with small mass splittings between states,  the modelling of initial state radiation can affect the signal significance. This modelling is taken from \herwig{} without modification.

In the limit of light neutralinos, with the assumption that the coloured sparticles are directly produced and
decay directly to jets and $\ninoone$, the limits on the gluino and squark masses are
approximately \absolteGluinoMassLimitSqGlPlane~GeV and \absolteSquarkMassLimitSqGlPlane~\GeV{} respectively for squark or gluino masses below 2 TeV, rising to \degenerateMassLimitSqGlPlane~GeV if the squarks and gluinos are assumed to be
mass-degenerate. These limits remain essentially unchanged if the $\ninoone$ mass is raised as high as 200 GeV. In the case of a specific SUSY-breaking scenario, i.e. CMSSM/MSUGRA
with $\tan\beta = 10$, $A_0 = 0$, $\mu>0$, the limit on $m_{1/2}$ reaches 460 GeV for
low values of $m_0$, and equal mass squarks and gluinos are excluded below \degenerateMassLimitSugraPlane~GeV. The use of signal selections sensitive to larger jet multiplicities than in \cite{daCosta:2011qk}
has improved the ATLAS reach at large $m_0$.
The five signal regions are used to set limits on $\sigma_{new}=\sigma A  \epsilon$, for non-SM cross-sections ($\sigma$) for which ATLAS has an acceptance $A$ and a detection efficiency of $\epsilon$ \cite{hepdata}. The excluded values of $\sigma_{new}$ are
22 fb, 25 fb, 429 fb,    
27 fb    
and
17~fb, 
respectively, at the 95\% confidence level.

\section{Summary}

This Letter reports a search for new physics in final states containing
high-\pT{} jets, missing transverse momentum and no electrons or muons with $\ourpt >20$~GeV. Data recorded by the ATLAS experiment a the LHC, corresponding to an integrated luminosity of \ourintlumi~ have been used.
Good agreement is seen between the numbers of events observed in the
five signal regions and the numbers of events expected from SM sources. 
The exclusion limits placed on non-SM cross sections impose new constraints on scenarios with novel physics.

The results are interpreted in both a simplified model containing
only squarks of the first two generations, a gluino octet and a
massless neutralino, as well as in MSUGRA/CMSSM models with $\tan\beta=10$, $A_0=0$
and $\mu>0$. In the simplified model, gluino and squark masses below \absolteGluinoMassLimitSqGlPlane~GeV and \absolteSquarkMassLimitSqGlPlane~\GeV{} respectively
are excluded at the $95\%$ confidence level for squark or gluino masses below 2 TeV, with the limit increasing
to  \degenerateMassLimitSqGlPlane~GeV for equal mass squarks and gluinos. In the MSUGRA/CMSSM
models, 
equal mass squarks and gluinos are excluded below \degenerateMassLimitSugraPlane~GeV.

\section{Acknowledgements}
We wish to thank CERN for the efficient commissioning and operation of the LHC during this data-taking period as well as the support staff from our institutions without whom ATLAS could not be operated efficiently.

We acknowledge the support of ANPCyT, Argentina; YerPhI, Armenia; ARC, Australia; BMWF, Austria; ANAS, Azerbaijan; SSTC, Belarus; CNPq and FAPESP, Brazil; NSERC, NRC and CFI, Canada; CERN; CONICYT, Chile; CAS, MOST and NSFC, China; COLCIENCIAS, Colombia; MSMT CR, MPO CR and VSC CR, Czech Republic; DNRF, DNSRC and Lundbeck Foundation, Denmark; ARTEMIS, European Union; IN2P3-CNRS, CEA-DSM/IRFU, France; GNAS, Georgia; BMBF, DFG, HGF, MPG and AvH Foundation, Germany; GSRT, Greece; ISF, MINERVA, GIF, DIP and Benoziyo Center, Israel; INFN, Italy; MEXT and JSPS, Japan; CNRST, Morocco; FOM and NWO, Netherlands; RCN, Norway;  MNiSW, Poland; GRICES and FCT, Portugal;  MERYS (MECTS), Romania;  MES of Russia and ROSATOM, Russian Federation; JINR; MSTD, Serbia; MSSR, Slovakia; ARRS and MVZT, Slovenia; DST/NRF, South Africa; MICINN, Spain; SRC and Wallenberg Foundation, Sweden; SER,  SNSF and Cantons of Bern and Geneva, Switzerland;  NSC, Taiwan; TAEK, Turkey; STFC, the Royal Society and Leverhulme Trust, United Kingdom; DOE and NSF, United States of America.  

The crucial computing support from all WLCG partners is acknowledged gratefully, in particular from CERN and the ATLAS Tier-1 facilities at TRIUMF (Canada), NDGF (Denmark, Norway, Sweden), CC-IN2P3 (France), KIT/GridKA (Germany), INFN-CNAF (Italy), NL-T1 (Netherlands), PIC (Spain), ASGC (Taiwan), RAL (UK) and BNL (USA) and in the Tier-2 facilities worldwide.

\vspace{5mm}


\bibliographystyle{atlasnote}
\bibliography{CONF_0lepton2011}

\clearpage
\onecolumn
\input{atlas_authlist}

\end{document}

%% file: table2.tex
\begin{table*}
\small
\begin{center}\renewcommand\arraystretch{1.4}
\begin{tabular}{|c|c|c|c|c|c|}
\hline\hline
\multirow{2}{*}{Process}     &\multicolumn{5}{|c|}{Signal Region} \\ 
\cline{2-6}
& \multirow{2}{*}{$\ge 2$-jet} & \multirow{2}{*}{$\ge 3$-jet} & $\ge 4$-jet,  & $\ge 4$-jet,  & \multirow{2}{*}{High mass} \\ 
&&&$\meff>500$~GeV&$\meff>1000$~GeV&\\
\hline
\hline

$Z/\gamma$+jets  & $ 32.3\pm\ph2.6\pm\ph6.9$ & $ 25.5\pm\ph2.6\pm\ph4.9$ & $ 209\pm\ph9\pm\ph38$ & $ 16.2\pm\ph2.2\pm\ph3.7$ & $ \ph3.3\pm\ph1.0\pm\ph1.3$ \\
 $W$+jets &$ 26.4\pm\ph4.0\pm\ph6.7$ &$ 22.6\pm\ph3.5\pm\ph5.6$ &$ 349\pm30\pm122$ &$ 13.0\pm\ph2.2\pm\ph4.7$ &$ \ph2.1\pm\ph0.8\pm\ph1.1$ \\
$t\bar{t}$+ single top  &$ \ph3.4\pm\ph1.6\pm\ph1.6$ &$ \ph5.9\pm\ph2.0\pm\ph2.2$ &$ 425\pm39\pm\ph84$ &$ \ph4.0\pm\ph1.3\pm\ph2.0$ &$ \ph5.7\pm\ph1.8\pm\ph1.9$ \\
QCD multi-jet   &$ 0.22\pm0.06\pm0.24$ &$ 0.92\pm0.12\pm0.46$ &$ \ph34\pm\ph2\pm\ph29$ &$ 0.73\pm0.14\pm0.50$ &$ 2.10\pm0.37\pm0.82$ \\

\hline

Total  &$ 62.4\pm\ph4.4\pm\ph9.3$ &$ \ph54.9\pm\ph3.9\pm\ph7.1$ &$ 1015\pm41\pm144$ &$ 33.9\pm\ph2.9\pm\ph6.2$ &$ 13.1\pm\ph1.9\pm\ph2.5$ \\

\hline\hline

Data  & 58 & 59 & 1118 & 40 & 18 \\

  \hline\hline
 \end{tabular}
\caption{\label{tab:SRevc} Fitted background components in each SR, compared with the number of events observed in data. The $Z/\gamma$+jets background is constrained with control regions CR1a and CR1b, the QCD multi-jet, $W$ and top quark backgrounds by control regions CR2, CR3 and CR4, respectively. In each case the first (second) quoted uncertainty is statistical (systematic). Background components are partially correlated and hence the uncertainties (statistical and systematic) on the total background estimates do not equal the quadrature sums of the uncertainties on the components. }
 \end{center}
 \end{table*}

%% file: table3.tex
\begin{table*}
\small
\begin{center}\renewcommand\arraystretch{1.4}
\begin{tabular}{|l|c|c|c|c|c|c|}
\hline\hline
\multirow{2}{*}{}     &\multicolumn{6}{|c|}{ Signal / Control Region} \\ 
\cline{2-7}
& CR1a & CR1b & CR2 & CR3 & CR4 & SR \\ 
\hline
\hline
Data  &8 &7 &34 & 15&12 & 18 \\ \hline
Targeted background  & $Z/\gamma$+jets& $Z/\gamma$+jets &  QCD multi-jet & $W$+jets &$t\bar{t}$ + single top &-- \\ \hline
Transfer factor &0.374 &0.812 &0.063 & 0.196&0.372 &--  \\ \hline
Fitted $Z/\gamma$+jets  &8.3 &5.8 &0.7 & 0.5&0.0& 3.3 \\
Fitted QCD multi-jet    &-- &-- &29.8 &0.8&0.6 & 2.1 \\ 
Fitted $W$+jets &-- &-- &0.5 & 10.0&0.4 & 2.1 \\
Fitted $t\bar{t}$ + single top  &-- &0.0 &3.0 & 3.7&11.0 & 5.7 \\ \hline
Fitted total background  &8.3 &5.9  &34.0  &15.0&12.0   & 13.1 \\
Statistical uncertainty   & $\pm$2.7 &$\pm$1.2 &$\pm$5.8 &$\pm$3.9 &$\pm$3.5 &$\pm$1.9 \\
Systematic uncertainty  & $\pm$0.6 &$\pm$1.7 &$\pm$0.1 &$\pm$0.1 &$\pm$0.2 & $\pm$2.5 \\
  \hline\hline
 \end{tabular}
\caption{\label{tab:SREdetails} Numerical inputs (i.e. the observed numbers of events in data) to and outputs from the likelihood fit to the control regions for the high mass channel. Each background process listed in the second row is measured using a control region, and the corresponding transfer factor is listed in the third row. An entry `--' in rows 5--7 indicates that the process in that row is assumed not to contribute to the control region (based on Monte Carlo studies) and hence is excluded from the fit. All numerical entries give event counts, with the exception of the transfer factors.}
 \end{center}
 \end{table*}

%% file: atlas_authlist.tex
\begin{flushleft}
{\Large The ATLAS Collaboration}

\bigskip

G.~Aad$^{\rm 48}$,
B.~Abbott$^{\rm 111}$,
J.~Abdallah$^{\rm 11}$,
A.A.~Abdelalim$^{\rm 49}$,
A.~Abdesselam$^{\rm 118}$,
O.~Abdinov$^{\rm 10}$,
B.~Abi$^{\rm 112}$,
M.~Abolins$^{\rm 88}$,
H.~Abramowicz$^{\rm 153}$,
H.~Abreu$^{\rm 115}$
% ...
\bigskip

% ...
$^{10}$ Institute of Physics, Azerbaijan Academy of Sciences, Baku, Azerbaijan\\
$^{11}$ Institut de F\'isica d'Altes Energies and Departament de F\'isica de la Universitat Aut\`onoma  de Barcelona and ICREA, Barcelona, Spain\\
% ...
$^{48}$ Fakult\"{a}t f\"{u}r Mathematik und Physik, Albert-Ludwigs-Universit\"{a}t, Freiburg i.Br., Germany\\
$^{49}$ Section de Physique, Universit\'e de Gen\`eve, Geneva, Switzerland\\
% ...
$^{88}$ Department of Physics and Astronomy, Michigan State University, East Lansing MI, United States of America\\
% ...
$^{111}$ Homer L. Dodge Department of Physics and Astronomy, University of Oklahoma, Norman OK, United States of America\\
$^{112}$ Department of Physics, Oklahoma State University, Stillwater OK, United States of America\\
% ...
$^{115}$ LAL, Univ. Paris-Sud and CNRS/IN2P3, Orsay, France\\
% ...
$^{118}$ Department of Physics, Oxford University, Oxford, United Kingdom\\
% ...
$^{153}$ Raymond and Beverly Sackler School of Physics and Astronomy, Tel Aviv University, Tel Aviv, Israel\\
% ...

\end{flushleft}
